\documentclass[11pt,twoside]{article}
\usepackage{asp2010}

\resetcounters

\begin{document}

\title{AstroCloud, a Cyber-Infrastructure for Astronomy Research: Architecture}
\author{Jian~Xiao$^1$, Ce~Yu$^1$, Chenzhou~Cui$^2$, Boliang~He$^2$, Changhua~Li$^2$, Dongwei~Fan$^2$, Zhi~Hong$^1$, Shucheng~Yin$^1$, Chuanjun~Wang$^3$, Zihuang~Cao$^2$, Yufeng~Fan$^3$, Shanshan~Li$^2$, Linying~Mi$^2$, Wanghui~Wan$^2$, Jianguo~Wang$^3$, Hailong~Zhang$^4$
\affil{$^1$Tianjin University, 92 Weijin Road, Tianjin 300072, China}
\affil{$^2$National Astronomical Observatories, Chinese Academy of Sciences (CAS), 20A Datun Road, Beijing 100012, China}
\affil{$^3$Yunnan Astronomical Observatory, CAS, P.0.Box110, Kunming 650011, China}
\affil{$^4$Xinjiang Astronomical Observatory, CAS, 150 Science 1-Street, Urumqi, Xinjiang 830011, China}
}

\begin{abstract}
AstroCloud is a cyber-Infrastructure for Astronomy Research initiated by Chinese Virtual Observatory (China-VO) under funding support from NDRC (National Development and Reform commission) and CAS (Chinese Academy of Sciences). The ultimate goal of this project is to provide a comprehensive end-to-end astronomy research environment where several independent systems seamlessly collaborate to support the full lifecycle of the modern observational astronomy based on big data, from proposal submission, to data archiving, data release, and to in-situ data analysis and processing. In this paper, the architecture and key designs of the AstroCloud platform are introduced, including data access middleware, access control and security framework, extendible proposal workflow, and system integration mechanism.
\end{abstract}

\section{Overview of the AstroCloud Platform}
\begin{figure}
  \centering
  \includegraphics[width=.8\textwidth]{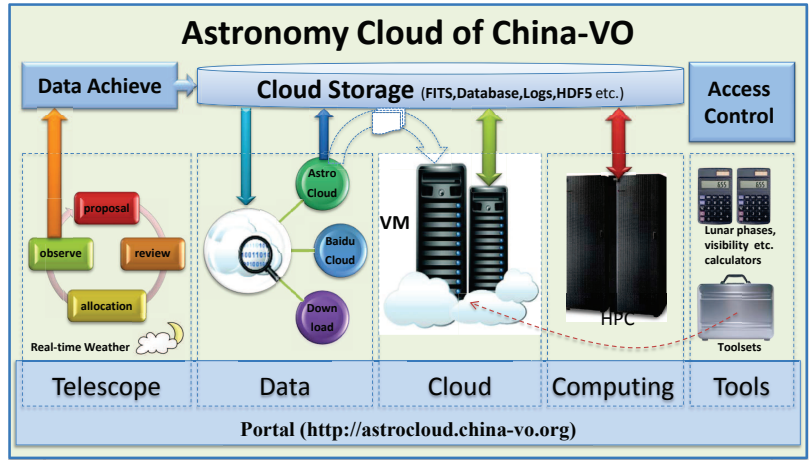}\\
  \caption{The overview of the Astrocloud, a platform built for full lifecycle of observation data management and in-situ data processing based on Cloud Computing technology}\label{fig:overview}
\end{figure}

Modern large digital telescopes have pushed astronomy into the data-driven era. Astronomers have to apply massive storage resources and computing power to process and analyze big data to probe the origins of our universe. Additionally, it becomes impractical to replicate copies at the sites of individual research groups (telescopes) ~\citep{bryant_2008}. In order to support the full lifecycle of the modern observational astronomy research based on big data, the AstroCloud platform was built by Chinese Virtual Observatory (China-VO), and as illustrated in Fig.~\ref{fig:overview}, the cyber-infrastructure consists of four main independent systems, which are Telescope, Data, Cloud, Computing(HPC), one auxiliary tools system, and two fundamental backend components for data archiving and storage. Except for the Telescope and Tools systems, other systems and components are built upon existing storage and computing resources located in five distributedly astronomical observatories of CAS. 

These independent systems seamlessly integrated to a comprehensive end-to-end astronomy research environment where astronomers can make proposal submission in the Telescope system, manage the observed data through the Data system, and perform data processing and analysis using VMs (virtual machine) in the Cloud system or more powerful physical clusters in the Computing system. In backend, the observed data from telescope can be automatically transferred to nearest local data center which is a node of the AstroCloud platform, so users can process their data in an in-situ manner within the platform which avoids large volume data transferring. 

Motivated by the integrated data-centric approach of Astro-WISE project~\citep{begeman_2013}, the AstroCloud platform adopted a similar data-driven architecture. As Fig.~\ref{fig:overview} demonstrated, the whole platform is built upon a distributed cloud storage layer where overall data are resident, in order to reduce the complexity of data management and performance optimization, a distributed data access middleware is designed to provide a unified data access interface. Upper systems can transparently access data without caring the physical location of the data. Moreover, based on the RBAC (role based access control) method and rule engine, a security framework is developed to provide unified user management, guarantee normal data access and protect private data. The platform is mainly implemented by Java programming language, plus a few of C/C++ codes, Python and Bash scripts.
\section{The Big-data Oriented Architecture}

\begin{figure}[!htbp]
  \centering
  \includegraphics[width=.8\textwidth]{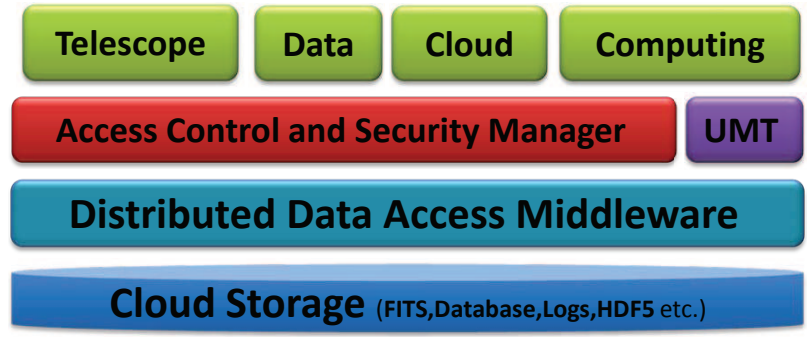}\\
  \caption{The overall architecture of the Astrocloud platform}\label{fig:arch}
\end{figure}

\subsection{Distributed Data Access Layer}
As shown in Fig.~\ref{fig:arch}, the dedicated cloud storage and unified data access layer are the fundamental backend of the whole AstroCloud platform. The cloud storage is based on GlusterFS file system and PostgreSQL database which only meta-data and star catalogues are stored in. Due to the large data volume, most data sets are distributed in many database partitions, moreover upper systems usually need to link to different databases and read files from different data nodes. The data access middleware provides a unified and efficient query interface which hides the complexity of data layouts to the upper systems. Besides that, the layer is also responsible for most housekeeping tasks related to data maintenance, such as data copy, backup and master/slave storage switching etc.. Most importantly, the data layer has a proven capability in keeping good query performance and horizontal scalability as the data volume continuously increases.

\subsection{RBAC and Rule Based Security Framework}
\begin{figure}[!htbp]
  \centering
  \includegraphics[width=.8\textwidth]{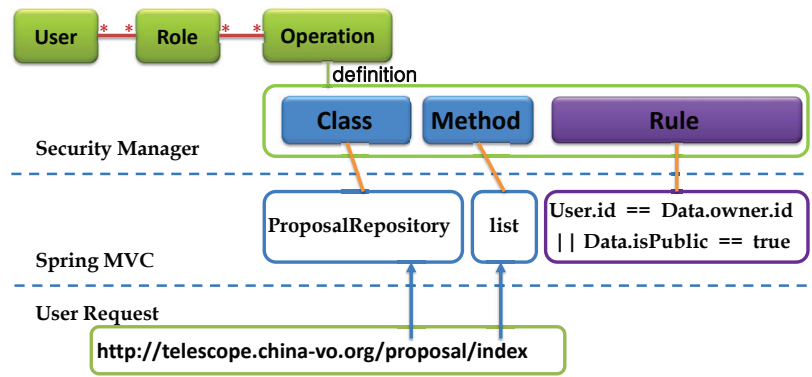}\\
  \caption{The RBAC and rule based security mechnism. RBAC is for coarse-grained check, and rule engine is for fine-grained control.}\label{fig:rbac}
\end{figure}

The security framework is implemented based on role-based access control(BRAC) method and rule engine. As illustrated in Fig.~\ref{fig:rbac}, user can have many roles, and role includes many permissions of operation, and operation is defined by the class and its method which implements this operation, plus a check rule. The rule describes the prerequisite for invoking this operation or the filter criteria for the underlying data. For example, user's request would be mapped to a particular java class and its some method, security manager will check whether the user has the permission to invoke the class's method, and then if passed, the rule engine will parse and evaluate the rule bound with the operation. According to the type of rule, some filtering constraints may be applied to the following execution for reading operation, or the authorization result will be returned for writing operation. 

\subsection{Extendible and Customizable Proposal Workflow}
\begin{figure}[!htbp]
  \centering
  \includegraphics[width=.8\textwidth]{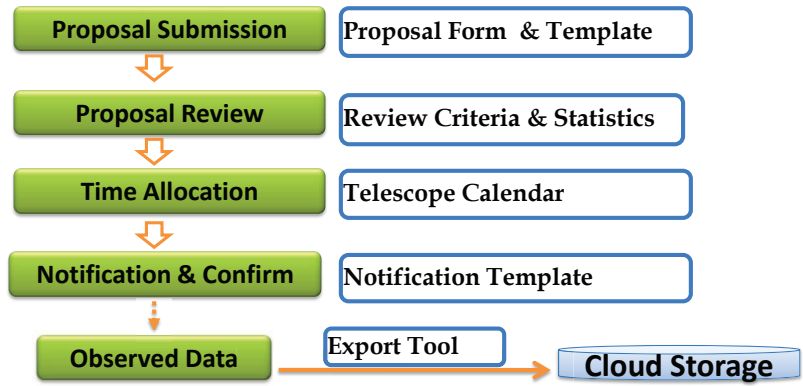}\\
  \caption{Telescope proposal management workflow. The unfilled rectangles represent its customizable points}\label{fig:workflow}
\end{figure}

In order to integrate new telescope quickly and support more telescopes in the future, an extendible proposal workflow is proposed. As illustrated in Fig.~\ref{fig:workflow}, the process of observation application is divided into four sequential phases, and the workflow engine will automatically drive the process. In each phase telescope-special customization is possible. For a telescope to participate in the platform, the developers only need to create necessary templates and write a few lines of code for telescope-special data model and its validation check. 
Usually telescope has its own dedicated control system and local limited storage, so it is necessary to implement a common data transfer interface---a bridge connecting individual telescopes and the platform, and a daemon process will periodically moving local observed data into the cloud storage through the bridge.

\section{Summary and Future Work}
At present the beta version of the AstroCloud platform have been released, and the 2.4m telescope on Lijiang station of Yunnan Observatories have been open in the platform. The integration tasks for the 2.16m telescope on Xinglong  station of National Observatories and the 25m radio telescope on Xinjiang Observatories are in progress.

\bibliographystyle{asp2010}	
\bibliography{P8-3}

\end{document}